\batchmode
\PassOptionsToPackage{utf8}{inputenc}
\PassOptionsToPackage{T1}{fontenc}
\documentclass[12pt,english]{article}
\usepackage{amsmath}
\usepackage{amsthm}

\usepackage{newtxmath}
\usepackage[T1]{fontenc}
\usepackage{color}
\usepackage{babel}
\usepackage{verbatim}
\usepackage{float}
\usepackage{graphicx}
\usepackage{geometry}
\usepackage{setspace}
\usepackage{ragged2e}
\usepackage[authoryear]{natbib}
\usepackage[colorlinks]{hyperref}  

\usepackage[utf8]{inputenc}
\usepackage[T1]{fontenc}

\hypersetup{
 linkcolor=red,urlcolor=red,citecolor=blue}

\makeatletter

\usepackage[bottom]{footmisc}
\usepackage{indentfirst}
\usepackage{lscape}
\usepackage{xcolor}
\usepackage{titlesec}
\date{February 2026}

\ifdefined\showcaptionsetup
 \PassOptionsToPackage{caption=false}{subfig}
\fi
\usepackage{subfig}
\makeatother

\begin{document}
\title{Cryptocurrency as an Investable Asset Class: \\ Coming of Age}
\author{Nicola Borri, Yukun Liu, Aleh Tsyvinski, and Xi Wu\thanks{Nicola Borri is with LUISS University, Rome. Yukun Liu is with the
University of Rochester, Simon Business School. Aleh Tsyvinski is
with Yale University. Xi Wu is with University of California, Berkeley. This review was prepared for \emph{Annual Review of Financial Economics}, Vol.18, 2026.}}
\maketitle
\begin{abstract}
\begin{singlespace}
\noindent We organize existing empirical regularities of cryptocurrencies into seven stylized facts and analyze cryptocurrencies through the lens of empirical asset pricing. We find important similarities with traditional markets---risk-adjusted performance so far is broadly comparable, and the cross-section of returns can be summarized by a small set of factors. However, cryptocurrency also has its own distinct character: jumps are frequent and large, and blockchain information helps drive prices. This common set of stylized facts provides evidence that cryptocurrency is emerging as an investable asset class. Additionally, we discuss potential data quality issues and possible changes in future regulations and the cryptocurrency environment.
\end{singlespace}
\end{abstract}
\thispagestyle{empty}

\pagebreak{}

\setcounter{page}{1}


\section{INTRODUCTION}

This survey establishes a set of seven key empirical regularities of cryptocurrencies motivated by the view that cryptocurrency is an asset and thus can and should be analyzed from the asset-pricing point of view and, more specifically, from the cryptocurrency asset pricing view.  Although the literature is relatively young---cryptocurrencies trace back to \citet{nakamoto2009bitcoin}'s short paper on Bitcoin---it is now at the point of maturity where it is useful to establish a core of stylized facts in cryptocurrency markets that can be used for both empirical and theoretical work. This common set of facts also provides emerging evidence that cryptocurrency is becoming an investable asset class.

Money has long been a central topic in economics---for example, see classic contributions of \citet{tobin1965money} and \citet{friedman1968role} to more recent advances. A large literature studied why money exists and how it functions---the role of money in economies with spatially separated agents (\citealt{townsend1980models}), the coexistence of different types of money under different institutional restrictions (\citealt{wallace1983legal}), search-based models \citep{kiyotaki1989money, trejos1995search,shi1995money,lagos2005unified}, mechanism-design approaches \citep{green1998rudimentary}, the ``money-as-memory'' view \citep{kocherlakota1998money}, commodity and token money \citep{sargent1983model,sargent2019commodity}, competing currencies \citep{kareken1981indeterminacy}, and the money-like safe assets (\citealt{gorton1990financial,holmstrom1998private}).

When the paper of \citet{nakamoto2009bitcoin} introduced a permissionless blockchain that provides public record-keeping without a central authority, financed and governed by a native token, the canonical literature on money had reached a degree of maturity. Permissionless ledgers can be viewed as technologies that supply public memory, with token systems financing and governing that record-keeping, and providing histories, communication, and enforcement mechanisms that are required for credit. With the rise of cryptocurrencies, these new ideas began to emerge primarily from practice more than theory and reopened classic monetary questions---record-keeping, commitment, enforcement---but now viewed from the point of view of blockchain economics. 

When practice advances faster than theory, it is natural to start asking empirical questions that can themselves later be used to shape theory. The paper of \citet{liu2021risks} is the first comprehensive analysis of cryptocurrency through the lens of empirical finance and asset pricing. A sizable literature uses an asset pricing lens to study risk exposures, cross‑sectional return patterns, cash‑flow‑like primitives, market microstructure, and limits to arbitrage (see, for example, \citealt{makarov2020trading}; \citealt*{liu2022common}; \citealt*{alvarez2023cryptocurrencies}; \citealt{sockin2023decentralization}; \citealt*{von2023decrypting}; \citealt{kogan2024cryptos}).

Many have viewed cryptocurrencies---and, not unjustifiably, many still view them---as hype, speculation, or fraud. When cryptocurrency first appeared it was easy to dismiss it: money engineered by an unknown author and operating outside of the traditional financial system.\footnote{See, \cite*{foley2019sex} and \cite{griffin2020bitcoin} for early analyses documenting how cryptocurrency mainly enabled illicit activity and alleged market manipulations.} It is worth noting that several important financial innovations, such as the limited‑liability company and mortgage‑backed securities, were also initially met with distrust before becoming part of standard finance (see, for example, \citealt{hilt2014corporate} and \citealt{lewis2010liar}), as institutions and legal frameworks evolved to manage their risks and as research deepened understanding of these innovations.

Since then research has evolved toward a much broader agenda that increasingly resembles research on any other asset class---the market structure (\citealt{abadi2018blockchain}; \citealt{leshno2020bitcoin}; \citealt{biais2019blockchain}; \citealt*{kogan2021economics}; \citealt*{brunnermeier2019digitalization}; \citealt*{benigno2022cryptocurrencies}; \citealt*{capponi2025maximal}; \citealt{jermann2025tokenomics}), regulation (\citealt{cong2019blockchain}; \citealt{townsend2020distributed}; \citealt*{auer2021distributed}), digital asset pricing and valuation (\citealt{athey2016bitcoin}; \citealt*{cong2021tokenomics}; \citealt{biais2023equilibrium}), transaction costs (\citealt*{easley2019mining}; \citealt*{cong2021decentralized}; \citealt{basu2023stablefees}), central bank money (\citealt{auer2022central}; \citealt*{whited2023will}; \citealt*{schilling2024central}), securitization (\citealt*{cong2022token}; \citealt{sockin2023decentralization}; \citealt{rogoff2023redeemable}; \citealt*{goldstein2024utility}), safe assets (\citealt*{ma2025stablecoin}), international capital flow (\citealt*{von2023decrypting}), blockchain technologies and the traditional financial system (\citealt{fernandez2018economics}) and market microstructure (\citealt{easley2024microstructure}).

We synthesize the rapidly growing empirical finance of cryptocurrencies into a set of seven stylized facts---empirical regularities that show how cryptocurrencies behave as assets, how their markets function, and what distinguishes them from traditional asset classes. These stylized facts help evaluate and revisit both classic and new theories of money, while remaining conditional on the available data and institutional environment to date and therefore potentially evolving as market structure, regulation, and adoption change.

\textbf{Seven Stylized Facts in Cryptocurrency Asset Pricing and Empirical Finance}. Stylized fact 1 overviews the core asset-pricing characteristics of cryptocurrency markets—the risk-return trade-off and the diversification benefits. Stylized fact 2 discusses the structure of returns in cryptocurrency—factors, or ``smart-beta'', which explain the cross-section of returns. Stylized fact 3 shows that the higher-order terms of a parsimonious four-factor model account for a large fraction of the cross-section of returns and provide a ``glass box'' alternative to the ``black box'' neural net models with interpretable economic factor structure. Stylized fact 4 shows the importance of the blockchain structure and network economy phenomena for cryptocurrency prices—on-chain user adoption influences cryptocurrency returns. Stylized fact 5 documents large and persistent market inefficiencies due to frictions that limit arbitrage, and discusses risks for arbitrageurs. Stylized fact 6 is about the cryptocurrency futures market and funding premia, and their evolution. Finally, stylized fact 7 is about the importance of regulation, disclosure, and transparency. In sum, these seven facts can be viewed as providing evidence that cryptocurrency, and more broadly, digital assets are coming of age as an emerging investable asset class.

\section{DATA}

The empirical analysis is based on the most recent data aggregated from multiple sources. 

First, we collect comprehensive price data on all cryptocurrencies from CoinGecko.com, covering both active and inactive coins between December 31, 2013, and September 6, 2025. The sample includes 18,622 coins listed on CoinGecko at the end of the period, as well as 29,230 inactive coins that have since been delisted but for which historical data are available. Including inactive coins allows us to avoid survivorship bias. We exclude stablecoins because they are designed to track the value of a currency, a basket of fiat currencies, or another reference asset. We also exclude wrapped tokens (such as Wrapped Ether) as they are essentially replicas of existing coins on different blockchains. Furthermore, we drop daily observations with missing or non-positive values for price, market capitalization, or trading volume. We also exclude all coins with fewer than 30 daily observations. The final dataset comprises 16,468 unique coins and approximately 9.5 million records. 

Second, for the construction of the crypto carry strategy of \cite*{schmeling2023crypto}, we collect perpetual futures prices, funding rates and spot index prices for Bitcoin at the 8-hour frequency from Binance for the period August 1, 2020 to May 31, 2025. The data on new addresses are from \cite*{liu2021accounting}. Moreover, for the construction of the crypto arbitrage strategy of \cite{borri2022cross}, we collect Bitcoin prices against 510 pairs from 165 different centralized exchanges quoted against 49 fiat currencies from CryptoCompare.com for the period December 31, 2013 to January 30, 2025. We convert these quotes into U.S.\ dollars using spot FX rates from Bloomberg. Furthermore, for estimation of jumps in cryptocurrency we collect tick-level transaction data for Bitcoin and Ethereum from Kraken.com and measures of realized variance constructed from tick-level data for the components of the Dow Jones Industrial Average Index from the CaPiRe dataset, for the period January 1, 2015 to March 31, 2025.\footnote{Finally, we obtain price indices for the MSCI global stock market, global high-yield and government bonds, a broad commodity index, gold, the U.S. dollar index, the Dow Jones US Real Estate Index, the U.S. constant maturity 10-year nominal and real bond yields from Bloomberg. The corresponding Bloomberg tickers are MXWO Index, LG30TRUU Index, LGTRTRUU Index, BCOM Index, XAU Curncy, DXY Curncy, DJUSRE Index, USGG10YR Index and GTII10 Govt.}

Unless otherwise noted, we aggregate daily closing prices to construct weekly coin returns.\footnote{CoinGecko daily prices are timestamped at 00{:}00 UTC (19{:}00 ET), whereas traditional asset prices from Bloomberg reflect end-of-day timestamps that vary by market and instrument (typically between 16{:}00 and 19{:}00 ET). Since our baseline analysis uses weekly close-to-close returns, any intraday timing mismatch is negligible. Crypto prices are quoted directly in USD, so no additional currency conversion is required. This review follows the dominant practice of using USD-denominated prices, which reflects the phenomenon that cryptocurrency trading is largely dominated by USD-linked quote currencies. We note that alternative numeraire or crypto-to-crypto return measures may reveal additional dimensions that are not captured by USD-based trades, which is a potentially important direction for future research.}
Each year is divided into 52 weeks: the first week contains the year’s first seven days, the next 50 weeks consist of seven days each, and the final week contains the remaining days of the year. For the construction of weekly returns, we exclude coins with a market capitalization below \$1 million at the formation date and winsorize returns at the 0.0025\% level in both tails. We construct a cryptocurrency market return as the value-weighted return of all underlying coins. The cryptocurrency excess market return is constructed as the difference between the cryptocurrency market return and the risk-free rate measured as the 1-month T-bill rate at the weekly frequency, which we obtain from the Federal Reserve Bank of St. Louis.

An important caveat concerns data integrity. A strand of the empirical cryptocurrency literature argues that parts of the historical crypto data---especially in early years---may reflect manipulation or unreliable reporting, including alleged price manipulation on Mt.\ Gox in 2013 \citep{gandal2018price}, inflated or ``artificial'' trading volume reported by some exchanges \citep{makarov2020trading}, and exchange-managed wash trading concentrated on ``unregulated'' exchanges and largely absent on ``regulated'' exchanges such as Bitstamp, Coinbase, and Gemini \citep{cong2023crypto}, as well as trade reordering/front-running on decentralized exchanges \citep{daian2019flash}. At the same time, on-chain evidence points to substantial cross-exchange flows: analyzing the Bitcoin blockchain, \cite{makarov2021blockchain} show that a sizable share of volume to exchanges originates from other exchanges or large market participants, consistent with arbitrage activity that tends to limit persistent price deviations across major venues. While these concerns cannot be fully eliminated, we mitigate them in several ways. First, our CoinGecko sample starts after the alleged Mt.\ Gox manipulation episode in 2013. Second, our pre-/post-2020 split places emphasis on the most recent period, and the post-2020 subsample begins after the main evidence documented in \cite{makarov2020trading} and \cite{cong2023crypto}. Third, the CoinGecko prices we use are volume-weighted averages across a curated set of large spot exchanges and, per CoinGecko's documentation,\footnote{See, \url{https://www.coingecko.com/en/methodology}.} are constructed using algorithms designed to detect and ignore outliers; as of the end of our sample these include the following exchanges: Binance, Coinbase, Kraken, Crypto.com, Bitfinex, Bitstamp, Gemini, and Huobi. Finally, our market-capitalization screens and data filters are designed to downweight or exclude small, thinly traded coins where data quality issues are more likely to be severe.

\begin{figure}[!htbp]
\centering
\subfloat[Return]{%
  \includegraphics[width=0.6\linewidth]{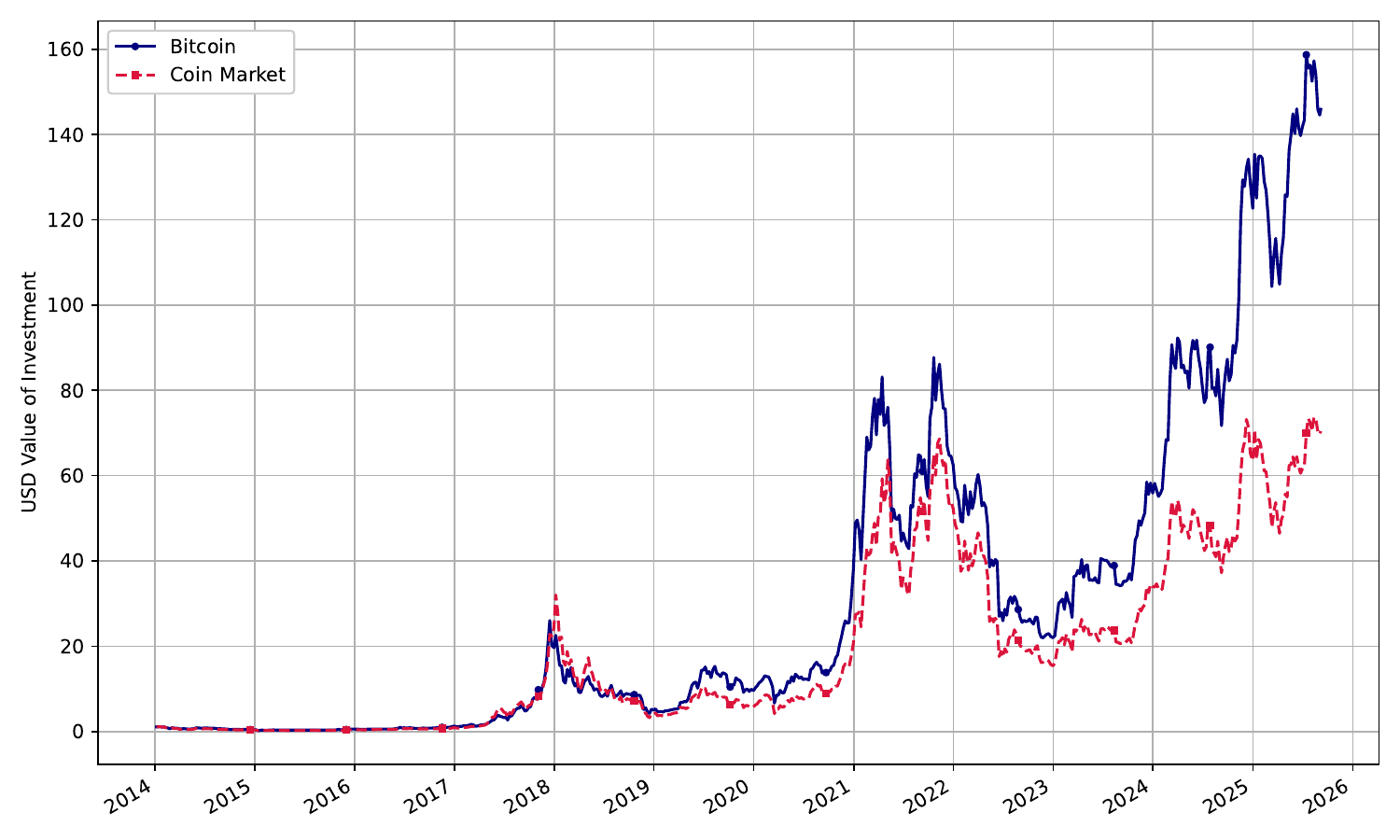}%
}\hfill
\subfloat[Capitalization]{%
  \includegraphics[width=0.6\linewidth]{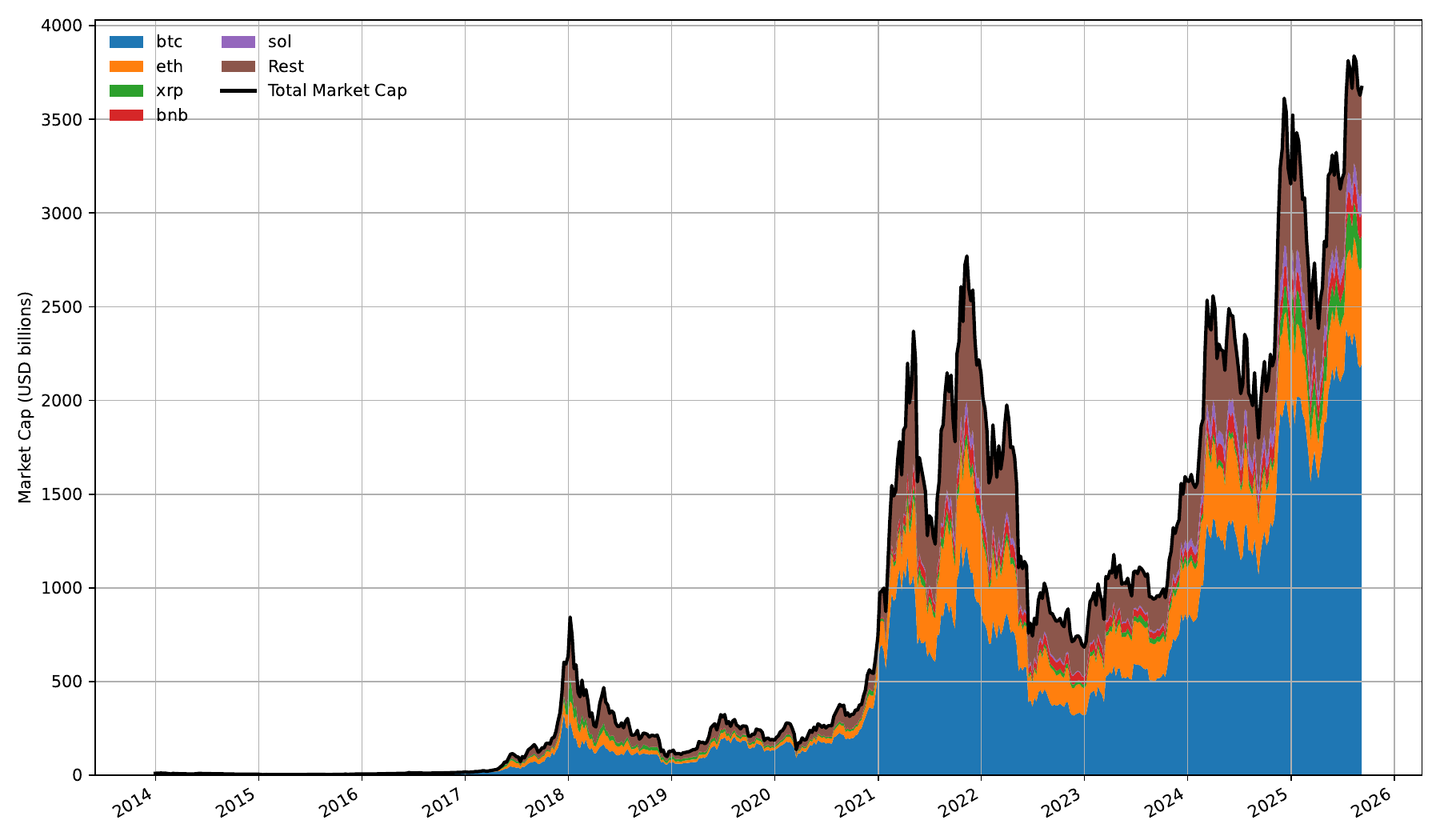}%
}
\caption{Cryptocurrency market returns and capitalization}
\label{fig:figure1}
\begin{minipage}{0.95\linewidth}
  \scriptsize
  \justifying
  \noindent Panel A of the figure plots the cryptocurrency market return against Bitcoin. Panel B plots the cryptocurrency market capitalization against the top five coins by market capitalization at the end of the sample (Bitcoin, Ethereum, Ripple, Binance Coin, and Solana). For returns, the figure shows the value of investment over time for one dollar of investment starting in December 31, 2013. For market capitalization, the figure shows the value over time of the outstanding coins in US dollar billions. Data end on September 6, 2025.
\end{minipage}
\end{figure}

\textbf{Figure \ref{fig:figure1}} summarizes the time-series evolution of cryptocurrency prices and market capitalization through September 2025. A one-dollar investment at the start of the sample grows to about \$160 in Bitcoin and \$80 in the overall market. There is substantial variation over time. Market capitalization reaches about \$4 trillion, with Bitcoin and Ethereum accounting for approximately 58\% and 12\% of the total at the end of the sample.\footnote{Since the end of our sample, aggregate cryptocurrency market capitalization has declined; as of late January 2026 it is about \$2.8 trillion.}

\section{THE SEVEN STYLIZED FACTS OF CRYPTOCURRENCY}

We organize the main discussions of this review into seven key empirical regularities, or ``stylized facts,'' on cryptocurrency markets.

\subsection{Stylized fact 1: High return, high volatility---normal Sharpe ratio so far, but rising comovement with equity}

\medskip
\noindent \textit{Stylized fact 1: The cryptocurrency market has an order of magnitude higher returns and volatility than equities, but a broadly similar Sharpe ratio; since 2020, it has also become substantially more correlated with equities.}
\medskip

Popular discussion of cryptocurrencies tends to focus either on their dramatic average returns or on dramatic volatility. From the classic finance point of view, the relevant measure is, however, risk-adjusted performance---the Sharpe ratio, i.e., mean excess return per unit of volatility (\citealt{sharpe1966mutual}).

\cite{liu2021risks} establish that the aggregate cryptocurrency market exhibits returns and volatilities an order of magnitude higher than those of the aggregate stock market, but Sharpe ratios remain broadly comparable. Stylized fact 1 revisits and confirms this finding in a longer dataset through September 2025.\footnote{These performance measures are computed from price indices and thus reflect the return to the underlying asset, not the return to a particular investor after custody risk. Episodes such as exchange cybersecurity breaches or bankruptcies (e.g., Mt.\ Gox and FTX) can generate additional losses that are not reflected in spot price series.}

\textbf{Table \ref{tab:tab1_sumstats}}, Panel A, reports summary statistics for the sample beginning in 2014 and ending in September 2025, as well as for the samples pre-2020 and post-2020. We compare returns on the aggregate coin market, a large-cap index of the top 100 cryptocurrencies (Coin 100), Bitcoin, and the aggregate global equity market. Across all three cryptocurrency indices, average weekly returns exceed 1\% and volatilities are about an order of magnitude larger than for global equities; however, weekly Sharpe ratios are similar across cryptocurrencies and equities. Panel A also reports drawdowns and run-ups, which are substantially larger for cryptocurrencies than for equities. Importantly, drawdowns and return volatility for cryptocurrencies are smaller in the post-2020 sample than pre-2020, while the reverse holds for equities.

A complementary change concerns diversification. \textbf{Table \ref{tab:tab1_sumstats}}, Panel B, reports correlations between the coin market return and major asset classes. In the pre-2020 period, correlations with traditional assets are small and statistically insignificant, consistent with \cite{liu2021risks}. In the post-2020 period, correlations rise sharply---most notably with global equities, increasing from 1\% pre-2020 to 39\% post-2020---and the coin market also becomes significantly correlated with commodities (27\%), high-yield bonds (24\%), government bonds (13\%), and gold (18\%). We also document a significant post-2020 correlation with changes in expected inflation (31\%), measured as the first difference in the yield differential between nominal and inflation-indexed U.S. Treasury bonds with 10-year maturity. Since this measure incorporates an inflation risk premium, the result may reflect the same underlying risk component that drives the stronger comovement between cryptocurrencies and equities in recent years. 

The increase in correlation between the cryptocurrency market and traditional asset classes has been documented in recent studies, for example, \cite*{dong2023tracing} and \cite{jiang2025financialization}. Several mechanisms are proposed to explain this phenomenon, including monetary policy, retail investor trading, and financialization. The literature has yet to reach a consensus on the cause of this important shift in the return dynamics of the cryptocurrency market.\footnote{Results are similar using a broad U.S. stock market index such as the S\&P 500 (2\% pre-2020; 37\% post-2020), while correlations with U.S. government bonds (rather than global government bonds) remain insignificant in the post-2020 period.} 

These correlation patterns matter directly for portfolio choice. In mean--variance settings, an asset's contribution to diversification depends on its covariance with the rest of the portfolio rather than its standalone volatility \citep{markowitz1952portfolio}. Consistent with the portfolio results in \cite{liu2021risks} and related Bayesian approaches \citep{duchin2022cryptocurrency}, a simple Black--Litterman calibration using post-2020 moments still implies non-trivial but modest allocations to the aggregate crypto market---about 3.1\% for a representative retail portfolio and 5.5\% for an institutional portfolio with a broader opportunity set. We emphasize that these magnitudes are back-of-the-envelope and depend on strong assumptions about expected returns, correlations, and investability (liquidity, custody, and regulatory constraints), and should therefore be interpreted as illustrative upper bounds on diversification benefits rather than investment advice.

\begin{table}[!htbp]
\caption{Summary statistics}
\label{tab:tab1_sumstats}
\begin{singlespace}
\begin{center}
\scriptsize
\setlength{\tabcolsep}{4pt}
\resizebox{\linewidth}{!}{%
\begin{tabular}{@{}lccccccccc@{}}
\hline
\multicolumn{9}{@{}l}{Panel A. Summary statistics returns} \\ 
\hline 
 & Mean (\%) & SD (\%) & t-Stat & Sharpe & Skew & Kurt & Max DD (\%) & Max RU (\%) \\
  & (1) & (2) & (3) & (4) & (5) & (6) & (7) & (8) \\  
 & \multicolumn{8}{c}{Full sample} \\ 
Coin market & 1.23 & 10.27 & 2.95 & 0.12 & 0.15 & 1.89 & -89.82 & 227.74 \\
Coin 100 & 1.24 & 10.30 & 2.98 & 0.12 & 0.15 & 1.84 & -89.23 & 224.08 \\
Bitcoin & 1.33 & 10.12 & 3.24 & 0.13 & 0.33 & 2.16 & -83.62 & 208.94 \\
Stock & 0.17 & 2.10 & 2.06 & 0.07 & -0.74 & 4.79 & -28.04 & 16.00 \\
& \multicolumn{8}{c}{Pre-2020 sample} \\ 
Coin market & 1.20 & 11.37 & 1.87 & 0.10 & 0.25 & 1.39 & -89.82 & 227.74 \\
Coin 100 & 1.20 & 11.44 & 1.86 & 0.10 & 0.23 & 1.29 & -89.23 & 224.08 \\
Bitcoin & 1.35 & 11.28 & 2.12 & 0.12 & 0.31 & 1.46 & -83.62 & 208.94 \\
Stock & 0.13 & 1.74 & 1.29 & 0.06 & -0.59 & 1.69 & -18.57 & 10.61 \\
 & \multicolumn{8}{c}{Post-2020 sample} \\ 
Coin market & 1.26 & 9.00 & 2.41 & 0.13 & -0.07 & 2.34 & -77.57 & 83.62 \\
Coin 100 & 1.29 & 8.98 & 2.46 & 0.14 & -0.03 & 2.41 & -76.78 & 84.36 \\
Bitcoin & 1.30 & 8.75 & 2.56 & 0.14 & 0.32 & 3.03 & -74.94 & 95.10 \\
Stock & 0.23 & 2.42 & 1.60 & 0.07 & -0.80 & 4.76 & -28.04 & 16.00 \\
\hline
\multicolumn{9}{@{}l}{Panel B. Coin market return correlations} \\
\hline 
Sample & Stock & Corp & Govt & Comm & Gold & Dollar & Inflation \\ 
 & (1) & (2) & (3) & (4) & (5) & (6) & (7) \\ \\
Full & 0.20 & 0.14 & 0.09 & 0.14 & 0.12 & -0.06 & 0.17 \\
 & (4.93) & (3.51) & (2.10) & (3.60) & (2.87) & (-1.49) & (4.35) \\
Pre 2020 & 0.01 & 0.03 & 0.05 & 0.03 & 0.06 & -0.03 & 0.03 \\
 & (0.15) & (0.48) & (0.88) & (0.51) & (1.13) & (-0.48) & (0.46) \\
Post 2019 & 0.39 & 0.24 & 0.13 & 0.27 & 0.18 & -0.10 & 0.31 \\
 & (7.31) & (4.28) & (2.20) & (4.74) & (3.10) & (-1.80) & (5.66) \\
\hline
\end{tabular}
}
\end{center}
\begin{scriptsize}
Notes: This table summarizes the main properties of coin market return. Panel A reports weekly summary statistics of the coin market index and compares them with the value-weighted index of the top 100 coins by market capitalization (Coin 100), Bitcoin and global stock market returns. Reported measures include the mean, standard deviation, $t$-statistic, Sharpe ratio, skewness, kurtosis, maximum drawdown and run-up. Panel B presents pairwise correlation coefficients between coin market returns and returns on the global stock market, global high-yield bonds, global government bonds, a commodity index, gold, the dollar index, and the changes in inflation expectations. $t$-statistics are reported in parentheses. The full sample covers the period from January 7, 2014 to September 6, 2025. The pre-2020 sample is the sample before Jan 1, 2020, and the post-2020 sample is the sample since Jan 1, 2020.
\end{scriptsize}
\end{singlespace}
\end{table}

\subsection{Stylized fact 2: How to be ``smart'' in crypto---crypto-size, crypto-momentum and crypto-value}

\medskip
\noindent \textit{Stylized fact 2: A small number of smart beta strategies---based on crypto-size, crypto-momentum, and crypto-value---span the cross-section of cryptocurrency returns.} 
\medskip

A central question in asset pricing is whether systematic variation in expected returns can be explained by a parsimonious set of factors. In equities, substantial academic endeavor---most prominently by Eugene Fama and Kenneth French (e.g., \citealt{fama1992cross})---established this approach, and factor models play a core role in modern finance practice and theory from risk management of large banks to niche hedge fund strategies. For cryptocurrencies, \cite*{liu2022common} establish that a three factor model consisting of the aggregate cryptocurrency market, size, and momentum (C-3 factor model), captures a large fraction of the cross-section of expected returns. \cite*{liu2021accounting} show that adding a value factor (C-4 factor model) further strengthens the explanatory power. We revisit this evidence using our updated sample and show that a small number of predictors drive the cross-section of cryptocurrency returns over the longer sample and post-2020. 

We construct smart beta strategies by sorting coins into quintile portfolios based on three established significant predictors: market capitalization (CSIZE), past two-week returns (CMOM), and the price-to-new-address ratio (CVALUE).\footnote{Studies based on smaller samples of coins have also proposed additional predictors of cryptocurrency returns in the cross-section, including liquidity (e.g., \citealt*{bianchi2022trading}), market segmentation (e.g., \citealt{borri2022cross}), and higher-order moments (e.g., \citealt{borri2022crypto, borri2022jumps}).} The return to each predictor is measured as the long-short spread between the highest and lowest quintiles (Portfolio 5 minus Portfolio 1).

\textbf{Table \ref{tab:portfolio_strategies_main}}, Panels A and B, report average excess returns to the quintile portfolios and the corresponding long-short strategies for the full sample and the post-2020 subsample. All three predictors generate statistically significant long-short returns in the full sample. The average long-short returns based on size and value are significantly negative and the average long-short return based on momentum is significantly positive, consistent with \cite*{liu2022common}. Cryptocurrency value is the weekly quintile portfolio of each bracket based on the price-to-new address ratio. High bracket is the low-value coins and low bracket is the high-value coins. The sorting is done in week $t$ and the returns are the value-weighted average in the following week.

Importantly, these results remain robust in the post-2020 sample, the core set of smart beta strategies in the C-3 factor model and C-4 factor model continues to explain variation in cryptocurrency returns even in the more recent period.

The C-4 structure of the crypto-market thus parallels the equity market, where a small number of equity-market factors---most notably size, value, momentum, and profitability---dominate the cross-section of expected returns. For example, \cite*{asness2013value} present evidence of value and momentum in equities, bonds, currencies, and commodities. The C-4 factor model of \cite*{liu2022common} and \cite*{liu2021accounting} is the cryptocurrency counterpart of the classic 4-factor model of \cite{fama1992cross} and \cite{carhart1997persistence} for equities. It is important to note that the predictors in cryptocurrency markets are only partly analogous to those in equity markets. CSIZE and CMOM closely resemble their equity counterparts, although crypto momentum is measured over a much shorter horizon (previous two weeks, versus the standard 12-to-2 months in equities). In contrast, there is no direct analogue to book value (\citealt{shiller1981stock}) in cryptocurrencies, so we follow \cite*{liu2021accounting} and proxy CVALUE with the price-to-new-address ratio.\footnote{\citet*{liu2021accounting} show that the number of new addresses captures the network effects of cryptocurrencies and serves as a measure of their fundamental value.} 

Establishing the small factor structure of the cryptocurrency returns, in our view, is not only important from purely the empirical asset pricing point of view but also since it provides a set of stylized low dimensional features that capture the main insights of several important theoretical models of cryptocurrency (\citealt*{cong2021tokenomics}; \citealt{sockin2023decentralization,sockin2023model}). More broadly, this stylized low-dimensional structure is consistent with the broader multifactor view of asset pricing developed for equities, but the relevant priced risks in cryptocurrency include distinct, crypto-specific components---including network/adoption-related fundamentals.

\begin{table}[!htbp]
\caption{Cryptocurrency smart beta strategies}
\label{tab:portfolio_strategies_main}
\begin{singlespace}
\begin{center}
\setlength{\tabcolsep}{6pt}
\begin{tabular}{@{}l cccccc@{}}
\hline
\multicolumn{7}{@{}l}{Panel A: Full sample}\\
\hline
& \multicolumn{6}{@{}c}{Quintiles} \\
\cline{2-7}
 & 1 & 2 & 3 & 4 & 5 & 5--1 \\ 
CSIZE & \textbf{Low} &  &  &  & \textbf{High} &  \\
Mean & 0.035*** & 0.015* & 0.014* & 0.007 & 0.012*** & -0.023*** \\
$t$(Mean) & (3.27) & (1.88) & (1.76) & (0.99) & (2.59) & (-2.66) \\
CMOM & \textbf{Low} &  &  &  & \textbf{High} &  \\
Mean & -0.002 & 0.005 & 0.001 & 0.018*** & 0.024*** & 0.026*** \\
$t$(Mean) & (-0.28) & (0.68) & (0.21) & (2.84) & (3.19) & (3.89) \\
CVALUE & \textbf{Low} &  &  &  & \textbf{High} &  \\
Mean & 0.034*** & 0.026** & 0.018** & 0.014 & -0.000 & -0.035*** \\
$t$(Mean) & (2.82) & (2.24) & (2.00) & (1.61) & (-0.05) & (-3.50) \\ 
\hline
\multicolumn{7}{@{}l}{Panel B: Post-2020 sample} \\ 
\hline
& \multicolumn{6}{@{}c}{Quintiles} \\
\cline{2-7}
 & 1 & 2 & 3 & 4 & 5 & 5--1 \\ 
CSIZE & \textbf{Low} &  &  &  & \textbf{High} &  \\
Mean & 0.021** & 0.011 & 0.012 & 0.013 & 0.012** & -0.009* \\
$t$(Mean) & (2.53) & (1.29) & (1.33) & (1.42) & (2.11) & (-1.69) \\
CMOM & \textbf{Low} &  &  &  & \textbf{High} &  \\
Mean & 0.001 & 0.001 & 0.009 & 0.016** & 0.021** & 0.021*** \\
$t$(Mean) & (0.08) & (0.17) & (1.38) & (2.41) & (2.50) & (3.70) \\
CVALUE & \textbf{Low} &  &  &  & \textbf{High} &  \\
Mean & 0.030** & 0.021* & 0.017** & 0.012* & 0.007 & -0.023** \\
$t$(Mean) & (2.35) & (1.95) & (2.06) & (1.79) & (1.35) & (-2.32) \\ 
\hline
\end{tabular}
\end{center}
\begin{scriptsize}
Notes: This table reports the mean quintile portfolio returns and long-short portfolio return for crypto smart beta strategies. The smart beta predictors are market capitalization (CSIZE), past two-week return (CMOM), and price-to-new address ratio (CVALUE). The long-short portfolio is long portfolio 5 and short portfolio 1 for all predictors. The mean returns are the time-series averages of weekly value-weighted portfolio excess returns. The post-2020 sample is the sample since Jan 1, 2020. $t$-statistics in parentheses are Newey-West adjusted. *, **, and *** denote significance at the 10\%, 5\% and 1\% levels. 
\end{scriptsize}
\end{singlespace}
\end{table}

A related important point is that in addition to determining which betas are smart, it is equally as important to establish which betas from the factor zoo are not “smart”. Specifically, \cite*{liu2022common} and \cite*{liu2021accounting} show that---besides crypto-specific size, momentum, and value---most of the smart beta strategies from the canonical factor zoo in the equity market (\citealt*{FGX20}) either do not generate significant hedged long-short strategy returns or are subsumed by the C-4 factor model. In Table \ref{tab:portfolio_strategies_not_work}, we study long-short strategies based on these signals post-2020. We find that these strategies continue not to generate significant average returns in the recent sample.\footnote{In a recent study, \cite{fieberg2024non} examine 20,736 research designs for 43 sorting variables in the cryptocurrency market. They find that the most prominent cryptocurrency factors, such as size and momentum, remain consistently robust across various specifications, but the others do not.} These results show that the cross-section of cryptocurrency returns can be summarized by a small number of factors.

\begin{table}[H]
\caption{Cryptocurrency smart beta strategies that are not beta}
\label{tab:portfolio_strategies_not_work}
\begin{center}
\scalebox{0.80}{
\begin{tabular}{@{}l cccccc@{}}
\hline
\multicolumn{7}{@{}l}{Panel A: Full sample}\\
\hline
& \multicolumn{6}{@{}c}{Quintiles} \\
\cline{2-7}
 & 1 & 2 & 3 & 4 & 5 & 5--1 \\
MOM12 & \textbf{Low} &  &  &  & \textbf{High} &  \\
Mean & 0.015** & 0.011 & 0.011* & 0.011* & 0.018*** & 0.003 \\
$t$(Mean) & (2.11) & (1.52) & (1.73) & (1.84) & (2.83) & (0.47) \\
MOM24 & \textbf{Low} &  &  &  & \textbf{High} &  \\
Mean & 0.021*** & 0.018** & 0.013** & 0.016** & 0.011* & -0.009 \\
$t$(Mean) & (2.81) & (2.37) & (2.22) & (2.48) & (1.96) & (-1.50) \\
VOL & \textbf{Low} &  &  &  & \textbf{High} &  \\
Mean & 0.013*** & 0.022*** & 0.014** & 0.015* & 0.011 & -0.003 \\
$t$(Mean) & (2.93) & (2.86) & (2.27) & (1.95) & (1.36) & (-0.45) \\
VOLUME & \textbf{Low} &  &  &  & \textbf{High} &  \\
Mean & 0.043** & 0.012* & 0.011 & 0.013* & 0.013*** & -0.030 \\
$t$(Mean) & (2.17) & (1.66) & (1.51) & (1.84) & (2.77) & (-1.60) \\
CMKT BETA & \textbf{Low} &  &  &  & \textbf{High} &  \\
Mean & 0.009 & 0.012** & 0.013** & 0.016** & 0.016** & 0.007 \\
$t$(Mean) & (1.47) & (2.17) & (2.35) & (2.50) & (2.38) & (1.34) \\
MKT BETA & \textbf{Low} &  &  &  & \textbf{High} &  \\
Mean & 0.006 & 0.012** & 0.015*** & 0.018** & 0.017* & 0.011* \\
$t$(Mean) & (0.88) & (2.19) & (2.81) & (2.48) & (1.96) & (1.80) \\ 
\hline
\multicolumn{7}{@{}l}{Panel B: Post-2020 sample} \\ 
\hline
 & 1 & 2 & 3 & 4 & 5 & 5--1 \\
 MOM12 & \textbf{Low} &  &  &  & \textbf{High} &  \\
Mean & 0.009 & 0.009 & 0.011 & 0.011* & 0.016** & 0.006 \\
$t$(Mean) & (1.18) & (1.15) & (1.59) & (1.84) & (2.24) & (1.30) \\
MOM24 & \textbf{Low} &  &  &  & \textbf{High} &  \\
Mean & 0.014* & 0.011 & 0.011 & 0.013* & 0.011* & -0.004 \\
$t$(Mean) & (1.72) & (1.52) & (1.59) & (1.90) & (1.75) & (-0.66) \\
VOL & \textbf{Low} &  &  &  & \textbf{High} &  \\
Mean & 0.012** & 0.011 & 0.014* & 0.011 & 0.005 & -0.007 \\
$t$(Mean) & (2.23) & (1.61) & (1.70) & (1.26) & (0.59) & (-1.39) \\
VOLUME & \textbf{Low} &  &  &  & \textbf{High} &  \\
Mean & 0.047* & 0.004 & 0.017 & 0.012 & 0.012** & -0.035 \\
$t$(Mean) & (1.69) & (0.67) & (1.60) & (1.63) & (2.16) & (-1.35) \\
CMKT BETA & \textbf{Low} &  &  &  & \textbf{High} &  \\
Mean & 0.006 & 0.013** & 0.014** & 0.009 & 0.014* & 0.008 \\
$t$(Mean) & (1.43) & (2.02) & (2.06) & (1.37) & (1.86) & (1.64) \\
MKT BETA & \textbf{Low} &  &  &  & \textbf{High} &  \\
Mean & 0.004 & 0.012** & 0.013** & 0.011* & 0.011 & 0.007 \\
$t$(Mean) & (0.61) & (2.00) & (2.00) & (1.67) & (1.27) & (1.17) \\ 
\hline
\end{tabular}
}
\end{center}
\begin{scriptsize}
Notes: This table reports the mean quintile portfolio returns and long-short portfolio return for crypto smart beta strategies based on insignificant predictors. The predictors are past 12-week and 24-week return (MOM12 and MOM24), past 24-week return standard deviation (VOL), past 24-week dollar trading volume (VOLUME), coin market beta (CMKT BETA) and stock market beta (MKT BETA) estimated using past 24-week returns. The long-short portfolio is long portfolio 5 and short portfolio 1 for all predictors. The mean returns are the time-series averages of weekly value-weighted portfolio excess returns. The post-2020 sample is the sample since Jan 1, 2020. $t$-statistics in parentheses are Newey-West adjusted. *, **, and *** denote significance at the 10\%, 5\% and 1\% levels.
\end{scriptsize}
\end{table}

\subsection{Stylized fact 3: Mind the Jumps---few factors, higher orders: why less is more}

\medskip
\noindent \textit{Stylized fact 3: Cryptocurrency returns feature frequent jumps; in the cross-section, a parsimonious model augmented with a small number of higher-order terms captures most of the predictive content of black-box machine learning.}
\medskip 

Leibniz wrote that \emph{natura non facit saltus}---nature makes no jumps. In economics and finance, this often means that most changes occur gradually, with only occasional discontinuities. Indeed, jumps are rare in many asset classes, and this key insight leads \cite{black1973pricing} to model prices as Brownian motions. Cryptocurrencies are different: large, sudden price moves are common. \cite{borri2022jumps} show that jumps account for a sizable share of Bitcoin's return variation and that higher-order moments help explain its risk premium.\footnote{See early evidence from \cite*{scaillet2020high} similarly documenting very frequent jumps in Bitcoin.}

\textbf{Table \ref{tab:crypto_jumps}} reports the frequency and magnitude of statistically significant jumps in Bitcoin and Ethereum---relative to U.S.\ equities---using realized-variance and bipower-variation measures constructed from 5-minute data and the BNS test. Jumps occur far more frequently in cryptocurrencies than in the Dow Jones stocks, although conditional on a jump the share of variation attributable to the jump is broadly similar across assets. In short, contrary to Leibniz's dictum, in crypto markets \emph{saltus sunt regula} (jumps are the rule), not the exception.

\begin{table}[ht]
\tabcolsep7.5pt
\caption{Jumps in Cryptocurrency}
\label{tab:crypto_jumps}
\begin{center}
\begin{tabular}{lcccccc}
\hline
 & Pr(J) & N(J) & \multicolumn{4}{c}{Quantiles of J/RV} \\
 & & & q(50) & q(75) & q(90) & q(95) \\ \hline

\multicolumn{7}{@{}c}{Full sample}\\ 

Bitcoin & 28.005 & 946 & 31.237 & 44.322 & 58.819 & 67.419 \\
Ethereum & 25.992 & 878 & 34.420 & 56.584 & 74.992 & 85.270 \\
Stock & 5.313 & 137 & 37.743 & 44.271 & 52.009 & 57.310 \\ 

\multicolumn{7}{@{}c}{Pre-2020 sample}\\ 

Bitcoin & 40.041 & 585 & 36.410 & 51.457 & 64.727 & 71.562 \\
Ethereum & 39.699 & 580 & 47.651 & 66.449 & 80.682 & 90.073 \\
Stock & 5.244 & 66 & 37.872 & 43.985 & 51.658 & 55.976 \\ 

\multicolumn{7}{@{}c}{Post-2020 sample}\\ 

Bitcoin & 18.841 & 361 & 26.161 & 33.197 & 42.350 & 48.892 \\
Ethereum & 15.501 & 297 & 24.545 & 30.736 & 39.429 & 45.667 \\
Stock & 5.643 & 60 & 37.959 & 44.838 & 52.434 & 57.079 \\ 
\hline
\end{tabular}
\end{center}
\begin{scriptsize}
This table reports the probability of observing a statistically significant jump (P(J)), the number of such events (N(J)), and quantiles of the ratio of jump variation to realized variance (J/RV), conditional on a significant jump, for Bitcoin, Ethereum, and U.S. stocks. Jumps are identified as the difference between the daily realized variance of \cite{barndorff2002econometric} and the bi-power variation of \cite{barndorff2006econometrics}. The two realized measures are constructed from tick-level data aggregated at the 5-minute frequency. Significant jumps are identified using the BNS test with a 0.1\% significance level. For U.S. equities, values are cross-sectional averages across the constituents of the Dow Jones Industrial Average. Data are from January 1, 2015 to March 31, 2025. The pre-2020 sample is the sample ending Jan 1, 2020. The post-2020 sample is the sample since Jan 1, 2020.
\end{scriptsize}
\end{table}

This prevalence of jumps, and the corresponding role of higher-order moments, poses challenges for risk management and portfolio choice (e.g., \citealt{borri2019conditional}) and has a direct implication for asset pricing: nonlinearities matter. Motivated by this, a growing literature applies machine-learning methods---especially neural networks---to the cross-section of returns (e.g., \citealt*{gu2020empirical}). While neural networks have a universal approximation property (\citealt{cybenko1989approximation}; \citealt*{hornik1989multilayer}), their main drawback in finance is that they are ``black box'' models: they can fit complex patterns but make it difficult to isolate which economic features drive predicted returns and whether those relationships are stable when the environment changes.\footnote{For a review of machine learning in asset pricing, see \cite{nagel2021machine}.}

One alternative to neural nets is the Kolmogorov-Arnold (KA) representation theorem (\citealt{Kolmogorov1956on,Arnold1957on}), introduced in finance by \cite{borri2024one}.\footnote{\cite{borri2024one}, arXiv, April 11, 2024.} The theorem shows that a broad class of nonlinear functions can be \emph{exactly} represented as finite sums of univariate functions composed with a link function---a structure that motivates \emph{shallow}, low-dimensional architectures (see, e.g., \citealt{hecht1987kolmogorov,girosi1989representation,kuurkova1991kolmogorov}). In practice, \cite{borri2024one} implement this idea and construct a single nonlinear factor (the KA factor) that captures a sizable share of the cross-sectional variation in expected returns.\footnote{Related work includes \cite{schmidt2021kolmogorov} and \cite{horowitz2007rate}. Subsequent to \cite{borri2024one}, \cite{liu2024kan} popularize KAN in the machine-learning community, while \cite*{toscano2025kkans} discuss computational and scaling limitations.}

Another alternative to neural networks follows the traditional finance toolbox---approximate and estimate the stochastic discount factor via a Taylor expansion. However, the Taylor approximation to the higher order generates a very large set of factors, including all of the interactions of the higher-order terms in the Taylor expansion. \cite{borri2025forward} develop the properties of a forward-selection Fama-MacBeth (FS-FMB) method for selecting the higher-order factors and their interactions that are most important for the cross-section of returns by sequentially adding one factor or interaction to the selected set at a time and, at each step, choosing the term that maximizes the Fama–MacBeth (FMB) second-step cross-sectional R-squared. This approach is a fully ``glass box'' model in terms of determining the important factors.\footnote{Conceptually, these higher-order interaction terms capture conditional moments, as in systemic-risk measures such as CoVaR \citep{tobias2016covar} and marginal expected shortfall \citep{acharya2017measuring}. See \cite{bell2024glass} for an alternative ``glass box'' machine learning structure.}

\textbf{Table \ref{tab:FS-FMB}}, Panel A, reports second-step FMB results for the crypto-CAPM, or C-1, model and for the C-4 model of \cite*{liu2021accounting}. The test assets are the set of 50 portfolios formed using the smart beta strategies described in stylized fact 2 (see \textbf{Tables \ref{tab:portfolio_strategies_main}} and \textbf{\ref{tab:portfolio_strategies_not_work}}), and the CARB strategy described in stylized fact 5 (see \textbf{Table \ref{tab:portfolio_strategies_arb}}, Panel B). As with the classic CAPM, the C-1 model cannot account for the cross-section of average returns. By contrast, the C-4 model explains 47.3\% of the cross-sectional variation, and satisfies the restriction of a zero intercept. \textbf{Table \ref{tab:FS-FMB}}, Panel B, reports the result of a specification which augments the C-4 model with the single KA-factor based on the Kolmogorov-Arnold architecture of \cite{borri2024one}. One can think of this factor as a stand-in that captures a variety of different neural nets. The cross-sectional R-squared increases by approximately 25 pp, and the intercept is not statistically different from zero at standard levels. \textbf{Table \ref{tab:FS-FMB}}, Panel C, reports the results from the estimation of the FS-FMB (\citealt{borri2025forward}). At each step, we augment the C-4 by one additional higher-order factor ($h_{j}$) which maximizes the adjusted cross-sectional R-squared. Higher-order factors include the squared and cubed of the factors in the C-4 model, as well their pair-wise interactions up to order three. The procedure selects three higher-order factors. The cross-sectional R-squared of the FS-FMB procedure is 70\%, a value very similar to that of the model including the KA-factor. In all models, the estimate of the intercept is not statistically different from zero. 

These results show that the FS-FMB procedure recovers essentially the same nonlinear structure as the KA single factor. Moreover, since KA can represent any neural net, FS–FMB provides a simple, transparent, ``open box'' alternative to black-box models that is directly connected to the economic structure of the model and to observables characteristics. In short, rather than estimating a neural network (or studying a particular variant of the multitude of neural networks), one can estimate the FS–FMB specification and capture the relevant nonlinearities in a parsimonious and interpretable way without relying on millions of parameters.

\begin{table}[ht]
\caption{Cross-Sectional Performance}\label{tab:FS-FMB}
\begin{center}
\begin{tabular}{llccc}
\hline 
\multicolumn{5}{c}{Panel A: Baseline Models}\\ 
\# & Model & Adj. R-squared & $\alpha$ & t-stat ($\alpha$) \\
\hline 
1 & C-1 & 0.046 & 0.012 & 1.437\\
2 & C-4 & 0.473 & 0.005 & 0.766\\ \\

\multicolumn{5}{c}{Panel B: KA model}\\ 

\# & Model & Adj. R-squared & $\alpha$ & t-stat ($\alpha$)\\
\hline
1 & C-4 \& KA & 0.723 & 0.011 & 1.494\\ \\

\multicolumn{5}{c}{Panel C: FS-FMB procedure}\\ 
\hline
Step & $h_{j}$ & Adj. R-squared & $\alpha$ & t-stat ($\alpha$)\\ 

1 & CMOM$^{3}$ & 0.623 & 0.002 & 0.235\\
2 & CMKT$^{2}$ & 0.663 & -0.001 & -0.140\\
3 & CMKT $\times$ CSIZE & 0.702 & 0.001 & 0.138\\
\hline
\end{tabular}
\end{center}
\begin{scriptsize}
This table reports the results from the second step of the FMB method.
Panel A refers to baseline models: the crypto-CAPM (C-1) and the C-4 model of \cite*{liu2022common}. The table reports the adjusted cross-sectional R-squared,
the estimate for the intercept ($\alpha$) and associated t-statistic.
The t-statistics are based on Newey-West corrected standard errors. Panel B refers to the C-4 model augmented by the single KA-factor, estimated with the KA procedure of \cite{borri2024one}.
Panel C refers to the FS-FMB procedure. At each step, we augment the
C-4 model by one additional higher-order factor ($h_{j}$) which maximizes
the adjusted cross-sectional R-squared. The method stops when the
improvement in the R-squared in a given step is smaller than 1 pp.
\end{scriptsize}
\end{table}

\subsection{Stylized fact 4: In crypto, the (block)chain drives the gain}

\medskip
\noindent \textit{Stylized fact 4: Blockchain economics and information significantly affect cryptocurrency prices, on-chain user adoption accounts for about 8\% of the variation in cryptocurrency returns.}
\medskip

Network externalities are a fundamental concept in economics because they shape markets and influence competition, regulation and technology adoption. Classic work has analyzed how networks form and evolve (\citealp{jackson1996strategic}), how pricing and competition operate on networked platforms (\citealp{rochet2003platform}), and how shocks propagate through financial networks (\citealp{allen2000financial}). A central question common in this work is information diffusion and learning in a network (\citealt{golub2010naive}; \citealt{acemoglu2011bayesian}).

One of the defining features of blockchain and one of the important innovations that it delivers is that blockchains create a continuously updated, public, and verifiable record of economic activity. Blockchains provide an observable, real-time map of user adoption and economic interactions. This makes them particularly useful for studying how information absorption affects asset values. As Metcalfe's law put it, ``the value of a network grows as the square of the number of its users.''

Empirically, \cite*{liu2021accounting} show that on-chain user adoption---measured as new user growth---has a statistically and economically significant effect on cryptocurrency prices. Importantly, this single on-chain variable accounts for about 8\% of the variation in cryptocurrency returns. In time‑series tests for the largest coins, on-chain user adoption alone can account for more than 10\% of the variation in their returns.

\cite*{liu2021accounting} also examine how markets absorb this information and what conditions amplify its relevance. Event‑time evidence shows a sharp price reaction when weekly adoption information is realized, with muted pre‑announcement behavior and no post‑announcement drift---in contrast to equities. The result gives evidence that investors fully incorporate the information at the weekly frequency.

Lastly, \cite*{liu2021accounting} show that the crypto value measure, constructed as the price-to-new-address ratio, negatively predicts cryptocurrency returns---a result we discuss in stylized fact 2. 

Taken together, these results connect to classic theories of networks and information diffusion and suggest that network-based adoption and propagation forces are quantitatively important for cryptocurrency returns---arguably more so than in mature equity markets.

\subsection{Stylized fact 5: Young cryptocurrency markets, old inefficiencies.}

\medskip
\noindent \textit{Stylized fact 5: The young cryptocurrency markets are still inefficient. But the origins are canonical: market segmentation and restrictions on capital flows. Cross-exchange cryptocurrency ``arbitrage'' is no free lunch.} 
\medskip

A central concept in finance is the law of one price---identical securities should sell for identical prices. The law of one price, and the related concept of absence of arbitrage, is important for the existence of a discount factor (\citealt{ross1978simple}; \citealt{harrison1979martingales}). Deviations from the law of one price have been presented as proof of irrational investor behavior (\citealt{lamont2003can}), limits to arbitrage and market segmentation (\citealt{shleifer1997limits}), or previously overlooked risks (\citealt{cochrane2002stocks}). 

One of the main features of cryptocurrency is that it is possible to trade coins simultaneously on multiple exchanges and against a number of fiat and crypto pairs. For example, an investor could buy Bitcoin on Binance or Kraken, two prominent centralized exchanges, against fiat currencies such as the U.S. dollar or the euro, or even against another cryptocurrency such as Ethereum or Tether, a stablecoin. 

\citet{makarov2020trading} and \citet{borri2023cryptomarket} document sizable, persistent deviations in the dollar price of Bitcoin across exchanges and across fiat- and crypto-quoted pairs. We revisit this evidence on updated data for 510 BTC–fiat pairs traded on 165 centralized exchanges in 49 currencies. Bitcoin price discounts are computed each week relative to the dollar price on Kraken (the benchmark exchange). All quotes are converted to U.S. dollars using daily spot FX rates. In the 2014–2025 sample, the 10th and 90th percentiles of Bitcoin price discounts are –1.8\% and 3.7\%, respectively. These figures remain virtually unchanged in the post-2020 period (–2.1\% and 3.5\%). Moreover, discounts are highly persistent, with a half-life of roughly 1.1 weeks in both samples. These deviations---classic examples of market inefficiency and violations of the law of one price---are results of market segmentation induced by capital controls and idiosyncratic domestic shocks that prevent arbitrage. 

Fragmentation and limits to arbitrage also have implications for market fragility and flash-crash dynamics. \citet{menkveld2019flash} study the May 6, 2010 U.S.\ equity Flash Crash and argue that a disruption in cross-market arbitrage can make fragmented markets fragile and result in price crashes; they also refer to analogous flash-crash episodes in cryptocurrency markets, including in Ethereum (June 21, 2017) and Bitcoin (October 10, 2017). Because our baseline analysis is conducted at the weekly frequency using close-to-close returns, these very short-lived intraday dislocations are less likely to affect our main results, though they remain relevant for studies of market microstructure and tail risk.\footnote{\citet{barbon2021quality} compare centralized and decentralized cryptocurrency exchanges along two key dimensions---market liquidity and price efficiency---and find that decentralized exchanges exhibit lower efficiency.}

\textbf{Table \ref{tab:portfolio_strategies_arb}}, Panel A, reports the properties of an ``arbitrage'' (CARB) strategy that attempts to exploit these price discrepancies. Each week, we sort BTC–fiat pairs by their discount and form five portfolios. The CARB return for a given pair is defined by buying Bitcoin on Kraken in week $t$ and selling it on the target pair in week $t{+}1$; portfolio returns are the simple average across constituent pairs. Using weekly formation and holding periods substantially alleviates execution and timing concerns. In practice, capital controls, withdrawal limits, transaction costs, settlement latency, and venue-specific frictions prevent the cross-market trades needed to enforce the law of one price. Not surprisingly, the mean excess returns are implausibly high—on the order of 68.5\% per week over 2014–2025, rising to 95\% per week from 2020 onward—precisely because the required arbitrage trades are blocked or prohibitively costly. Specifically, \cite{borri2022cross} show that this seemingly exceptional performance is concentrated in the extreme portfolio that holds pairs with the highest Bitcoin price relative to the benchmark quote, where segmentation and frictions are most severe. That is, the cryptocurrency market inefficiencies reflect the classic market segmentation and trading frictions as in \cite{shleifer1997limits}.

As Sophocles writes in \emph{Antigone}, ``Nothing vast enters the life of mortals without a curse.'' Early cryptocurrency investors often told a different story: that outsized returns came from cross-exchange ``arbitrage.'' Public interviews with Sam Bankman-Fried—the founder of the now-defunct exchange FTX—describe how Alameda Research, a hedge fund he helped run, allegedly began by focusing on such trades.\footnote{See, for example, Bloomberg's coverage of the so-called Kimchi premium, which denotes the fact that it was more expensive to buy Bitcoin in Korean Won than in US dollars: \url{https://www.bloomberg.com/news/articles/2024-11-15/south-korea-crypto-boom-what-is-the-kimchi-premium}, and \cite{lewis2023going}'s book about Sam Bankman-Fried.} However, \cite{borri2022cross} show that implementability is limited to a narrow set of highly liquid pairs on reputable cryptocurrency exchanges without binding capital controls. 

We revisit the analysis in \cite{borri2022cross} and construct the CARB strategy for 226 Bitcoin pairs, traded on 96 exchanges against 11 fiat currencies (Australian dollar, Canadian dollar, Danish krone, euro, Hong Kong dollar, Israeli shekel, Japanese yen, Polish złoty, Swiss franc, British pound, and U.S. dollar). These are the pairs that are realistically investable. \textbf{Table \ref{tab:portfolio_strategies_arb}}, Panel B, documents that excess returns remain statistically significant but are \emph{orders of magnitude smaller}: the long–short weekly return declines from about $0.68$ to roughly $0.038$ in the full sample, and measures around $0.041$ in the post-2020 sample. In other words, where the strategy is actually implementable, the apparent ``free lunch'' largely vanishes. These returns are not pure arbitrage---they are compensation for risk. \citet{borri2022cross} show that the expensive pairs (those priced above the benchmark) are systematically \emph{riskier}—they depreciate more in bad times for crypto-investors, when aggregate liquidity and sentiment are low. 

The collapse of Alameda Research and the failure of FTX in November 2022—only a few months after the demise of the synthetic stablecoin ecosystem Terra–Luna described in \cite*{liu2023anatomy}, is the most visible manifestation of the results described in this fact. More broadly, the evidence from cryptocurrency markets provides a clean illustration of the classic limits-to-arbitrage view: when trading, settlement, and capital-mobility frictions are economically large, violations of the law of one price can persist rather than being immediately arbitraged away.

\begin{table}[ht]
\tabcolsep7.5pt
\caption{Cryptocurrency ``Arbitrage'' Strategy}
\label{tab:portfolio_strategies_arb}
\begin{center}
\begin{tabular}{@{}l cccccc@{}}
\hline
\multicolumn{7}{@{}l}{Panel A: All exchanges}\\
\hline
& \multicolumn{6}{@{}c}{Quintiles} \\
\cline{2-7}
 & 1 & 2 & 3 & 4 & 5 & 5--1 \\ 
 
& \multicolumn{6}{@{}c}{Full sample} \\ 
CARB & \textbf{Low} &  &  &  & \textbf{High} &  \\
Mean & -0.011** & 0.009* & 0.012** & 0.017*** & 0.674*** & 0.685*** \\
$t$(Mean) & (-2.22) & (1.73) & (2.43) & (3.38) & (4.39) & (4.46) \\ 

& \multicolumn{6}{@{}c}{Post-2020 sample} \\ 
CARB & \textbf{Low} &  &  &  & \textbf{High} &  \\
Mean & -0.014** & 0.006 & 0.010* & 0.015*** & 0.940*** & 0.954*** \\
$t$(Mean) & (-2.43) & (1.01) & (1.76) & (2.75) & (3.90) & (3.97) \\ 
\hline
\multicolumn{7}{@{}l}{Panel B: Exchanges in open markets}\\
\hline
& \multicolumn{6}{@{}c}{Quintiles} \\
\cline{2-7}
 & 1 & 2 & 3 & 4 & 5 & 5--1 \\ 
& \multicolumn{6}{@{}c}{Full sample} \\ 
CARB & \textbf{Low} &  &  &  & \textbf{High} &  \\
Mean & -0.002 & 0.007 & 0.010* & 0.014*** & 0.036*** & 0.038*** \\
$t$(Mean) & (-0.39) & (1.36) & (1.94) & (2.71) & (7.66) & (12.45) \\ 
& \multicolumn{6}{@{}c}{Post-2020 sample} \\ 
CARB & \textbf{Low} &  &  &  & \textbf{High} &  \\
Mean & -0.008 & 0.005 & 0.008 & 0.012** & 0.033*** & 0.041*** \\
$t$(Mean) & (-1.36) & (0.75) & (1.26) & (2.13) & (6.06) & (9.90) \\ 
\hline
\end{tabular}
\end{center}
\begin{scriptsize}
Notes: This table reports the mean quintile portfolio returns and long-short portfolio return for the ``arbitrage'' (CARB) investment strategy that sorts Bitcoin trading pairs by their price deviation from a benchmark Bitcoin price on Kraken. Each week, pairs are sorted into five portfolios. The long-short portfolio is long portfolio 5 and short portfolio 1. The mean returns are the time-series averages of weekly equally-weighted portfolio excess returns. Panel A uses the full sample of 510 BTC–fiat pairs traded on 165 centralized exchanges in 49 currencies. Panel B restricts to 226 BTC–fiat pairs on 96 exchanges against 11 fiat currencies, which are the most liquid fiat currencies on the most reputable exchanges in jurisdictions with minimal capital controls as in \cite{borri2022cross}. The post-2020 sample is the sample since Jan 1, 2020.  $t$-statistics in parentheses are Newey-West adjusted. *, **, and *** denote significance at the 10\%, 5\% and 1\% levels. 
\end{scriptsize}
\end{table}

\subsection{Stylized fact 6: When the funding dries up, we finally learn the worth of futures}

\medskip
\noindent \textit{Stylized fact 6: Cryptocurrency carry trade, the defining characteristics of the futures market, used to be highly profitable, but profitability has compressed sharply since 2024.}
\medskip

Derivatives are a cornerstone of modern markets, enabling hedging, leverage, and price discovery across equities, interest rates, and commodities (\citealt{black1973pricing}; \citealt{b5405564-0209-3283-bc5e-72bd72cfbbed}; \citealt{black1976pricing}). 

There is widespread usage of derivatives in the cryptocurrency market, albeit with novel designs. Among all derivatives, perpetual futures are the most common derivative contracts in cryptocurrency markets. For example, \cite*{schmeling2023crypto} document that perpetual futures account for more than 98\% of the Bitcoin futures volume in their sample.  Perpetuals, inspired by \cite{shiller1993measuring}, have no maturity; positions are marked to market (typically every 8 hours) and a funding rate---positive or negative by design---flows between longs and shorts to keep futures prices anchored to spot. \cite*{ackerer2024perpetual} provides the theoretical foundations for the pricing of perpetual cryptocurrency futures. The crypto-carry trade strategy of \cite*{schmeling2023crypto} is short a perpetual futures contract and long a position in the corresponding spot market. 

\begin{figure}[h]
\begin{center}
\includegraphics[scale=0.42]{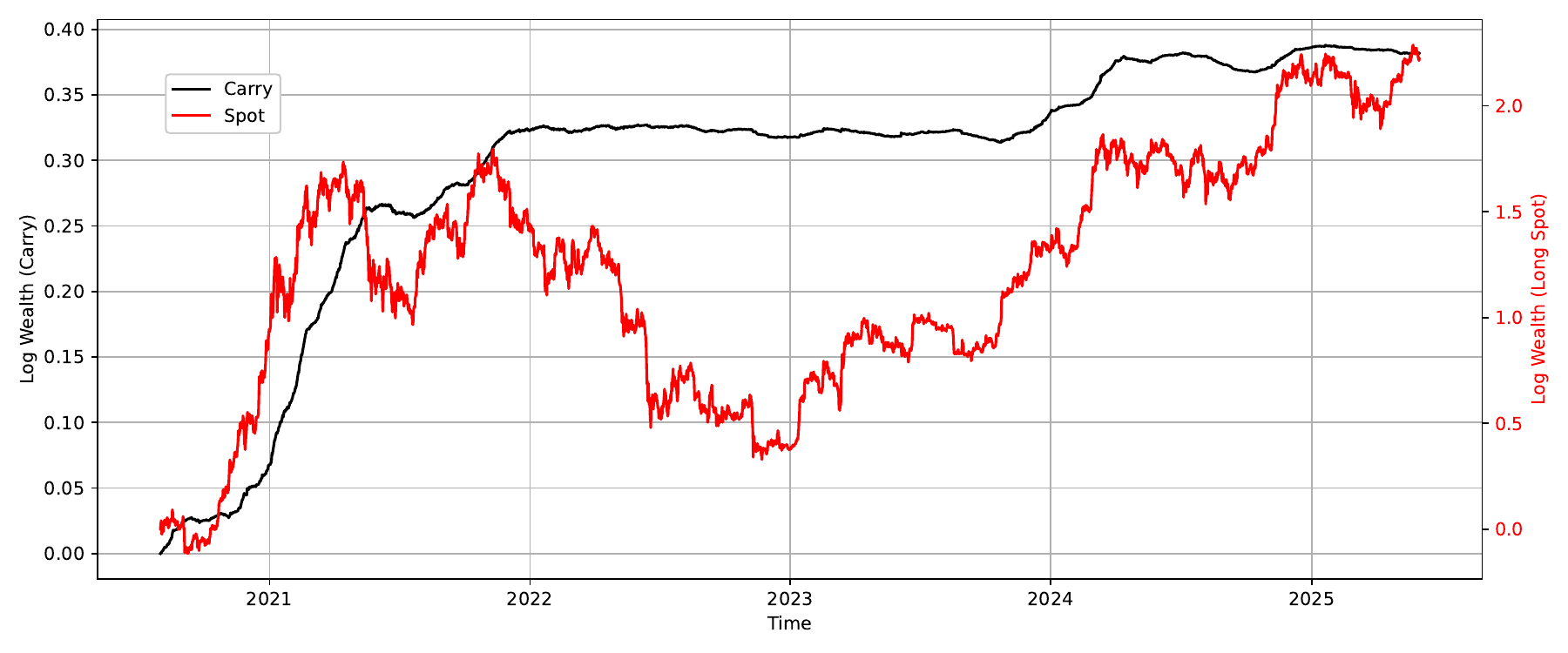}
\end{center}
\bigskip
\caption{Carry Trade Cumulative Returns}
\medskip
The figure plots the cumulative returns from Tether Carry Trades (left axis) and from a long buy-and-hold strategy in Bitcoin (right axis). The Carry Trade is defined as in \cite*{schmeling2023crypto}. The sample goes from August 1, 2020 to May 31, 2025.
\label{fig:figure_carry_futures}
\end{figure}

\textbf{Figure \ref{fig:figure_carry_futures}} plots the evolution over time of one dollar invested in the crypto-carry strategy (black line, left y-axis) and in the spot Bitcoin market (red line, right y-axis). Over the full sample, which goes from 2020 to 2025, the annualized Sharpe ratio of the cryptocurrency carry is 6.45. Beginning in 2024, the Sharpe ratio falls to 4.06, and it turns negative in 2025. The profit from the cryptocurrency carry strategy is mostly driven by the funding rate, which in the full sample has a mean return of approximately 8\% with a low volatility of 0.8\%. 

This shift in the profitability of the crypto-carry trade strategy has important implications for the crypto-industry. A growing set of crypto ``yield'' products---most prominently delta-neutral, funding-rate-harvesting structures (e.g., synthetic dollar projects that combine spot holdings with short perpetuals)---derive an important part of their returns from the funding rate. A prominent example is Ethena.fi, whose stablecoin uses a delta-neutral construction and relies on perpetual-funding payments. As of September 2025, Ethena's stablecoin reached a market capitalization of about \$14 billion, with its governance token ENA valued at around \$4 billion. If funding premia compress or become more volatile, the economics of these strategies deteriorate and the expected yields decline. The lower profitability of the crypto-carry trade shows that funding-rate premia are neither guaranteed nor permanent and, even more importantly, raises difficult questions about the long-run sustainability of yield products built on them.

\subsection{Stylized fact 7: Growing up with supervision---regulation and oversight strengthen cryptocurrency markets}

\medskip
\noindent \textit{Stylized fact 7: As the market matures, there is growing disclosure and transparency in cryptocurrencies.}
\medskip

A fundamental insight of information economics is that markets with asymmetric information can fail to allocate resources efficiently. When one party possesses superior information, such as inside information, prices may deviate from fundamentals, and capital may be misallocated (e.g., \citealt{akerlof1970market}). Accounting research builds directly on this and shows how disclosure, regulation, and intermediaries can reduce these frictions (\citealt{verrecchia1983discretionary}; \citealt{dye1985disclosure}; \citealt{healy2001information}; \citealt{verrecchia2001essays}). However, in the absence of regulation and supervision, managers may strategically disclose for private benefits (\citealt{verrecchia2001essays}; \citealt{ma2025invention}). Policymakers thus issue regulations to address these inefficiencies and to produce credible, comparable, and decision-useful information. Intermediaries, such as analysts and auditors, emerge to reduce information asymmetry by analyzing and validating information (\citealt{watts1983agency}).

Information asymmetry remains pervasive in cryptocurrency markets, though regulation and disclosure have increased in recent years. On the corporate side, public firms with material cryptocurrency exposure are now required to disclose their cryptocurrency holdings and risks. Importantly, the introduction of Accounting Standards Update (ASU) 2023-08 by the Financial Accounting Standards Board provides a coherent framework for the accounting of crypto-assets, which formally requires fair-value accounting treatment (\citealt{FASB2023CryptoAssets}). Recent accounting research emphasizes measurement reliability and disclosure consistency for digital assets (\citealt{anderson2025accounting}).

On the cryptocurrency side, voluntary disclosures, such as whitepapers and social media announcements, have become increasingly common (e.g., \citealt{george2021blockchain}; \citealt*{liu2025investors}; \citealt*{du2025voluntary}). \cite*{howell2020initial} and \cite{bourveau2022role} show the importance of voluntary disclosures in the initial coin offering stage. Studies also examine information intermediaries (e.g., analysts and auditors) and find that they play an important role in this market (e.g., \citealt{bourveau2022role};  \citealt*{bourveau2024decentralized}). 

In July 2025, Congress enacted the Guiding and Establishing National Innovation for U.S. Stablecoins Act (GENIUS Act), the first comprehensive federal framework for digital asset oversight in the United States. The Act focuses primarily on payment stablecoins, aiming to improve their transparency and stability. It mandates monthly public disclosures of reserve compositions of stablecoins, and explicitly prohibits stablecoin companies from making misleading claims that their tokens are backed by the U.S. government, federally insured, or legal tender. 

The recent regulations and market developments show that greater transparency in the cryptocurrency space is not only increasingly required by the regulators but is also valued by the market. The cryptocurrency market is starting to increasingly implement and value more rigorous accounting principles and slowly converging to the accounting practices of traditional asset classes, for which reliable, comparable, and verifiable disclosure is a major and perhaps the most important determinant of investor confidence and growth of an asset class.

\section{ADDITIONAL DISCUSSIONS}

The seven stylized facts summarize key developments in cryptocurrencies through the lens of empirical asset pricing. In this section, we provide additional discussions on emerging topics, including retail investor trading in the cryptocurrency market, the non-fungible token market, and studies on the dark side of cryptocurrency usage.

Recent research increasingly examines individual investor beliefs and behavior in cryptocurrency markets and compares and contrast them with those in traditional investments. \cite{kogan2024cryptos} show that the trading behaviors of retail investors are drastically different between cryptocurrencies and traditional assets, such as stocks and gold. In particular, retail investors are contrarian in stocks and gold, but trend-chasing in cryptocurrencies. \cite{kogan2024cryptos} conjecture that the trend-chasing behavior is due to cryptocurrency price changes affecting the likelihood of future widespread adoption, which leads retail investors to update their price expectations. \citet{chan2020inside} use data from a medium-sized cryptocurrency exchange and document that investor portfolios are typically small in value, highly concentrated in a few coins, and characterized by short holding periods. \citet{weber2023you} complement this evidence using repeated large-scale surveys of U.S. households, showing that cryptocurrency investors differ systematically from other investors. \citet*{alvarez2023cryptocurrencies} provide complementary evidence on the use of cryptocurrencies as a means of payment, using the case of El Salvador—where Bitcoin was made legal tender—and show that privacy and security concerns remain key barriers to adoption. \cite*{han2023social} study how social interactions and sentiments affect retail investor trading in cryptocurrency markets. Finally, \cite{benetton2024investors} incorporate individual-level survey data into a structural characteristics-based demand model where investors can hold cryptocurrencies. They find that individual beliefs of cryptocurrency are different and are critical in explaining cryptocurrency valuation.

The bulk of this review is about fungible tokens. A significant segment of the cryptocurrency market, however, is non-fungible tokens (NFTs). \cite{goetzmann2022non} are the first to construct repeated sales return indices of NFTs and examine their properties. \cite{borri2022economics} provide a comprehensive analysis of the NFT market. They highlight the importance of visual features in cryptocurrency valuation and how it leads to a conspicuous consumption effect in the market. Furthermore, \cite{borri2022economics} show that, due to the high non-fungibility of NFTs, the market is highly segmented, which has implications for returns and investor behaviors in the market. \citet{huang2023selection} document that behavioral biases---such as selection neglect and extrapolative beliefs---play a significant role in NFT pricing. \cite*{oh2025investor} document preference-driven herding in NFT primary market sales.

A stream of work examines the so-called dark side of cryptocurrencies, especially in the early years. \cite*{foley2019sex} argue that in the early days, the majority of Bitcoin transactions were about illegal activities, but this is recently challenged by \cite{makarov2021blockchain} that show the results of \cite*{foley2019sex} are largely driven by double-counting. \cite{griffin2020bitcoin} analyze blockchain and market data to argue that Tether flows timed after market downturns are associated with large Bitcoin price increases during the 2017 boom, consistent with price manipulation. \cite*{amiram2022coins} show that abnormal spikes in on‑chain transfers around major terrorist attacks carry predictive power for attack timing and highlight chokepoints visible in public ledgers, \cite{cong2025anatomy} provide an anatomy of crypto-enabled crimes and identify ransomware as the dominant activity, while \cite{cong2023dark} review crypto-related scams, including Ponzi schemes and ``rug pulls'' in decentralized projects and token offerings. \cite*{li2018cryptocurrency} document pervasive cryptocurrency pump‑and‑dump episodes that generate brief surges in prices, volume, and volatility followed by rapid reversals, with pre‑event run‑ups consistent with insider gains at outsiders’ expense.

\section{CONCLUSIONS}

In the 17 years since Nakamoto, cryptocurrencies have reached a certain stage of maturity: they can now be studied using the same measurement, data quality, and research rigor applied to other asset classes. By the end of our sample period (September 2025), aggregate cryptocurrency market capitalization is about \$4 trillion. Major financial firms are using cryptocurrency in their businesses.

We show that there are important similarities between cryptocurrency markets and traditional asset markets---Sharpe ratios so far are broadly comparable, and a small number of factors and their higher orders summarize the cross-section of returns. On the other hand, cryptocurrency also has its own character---frequent jumps and blockchain information driving prices. The cryptocurrency market has become more mature and is emerging as an asset class with its own distinct characteristics that still, however, needs more data, research, supervision, enforcement, and a better legal and regulatory framework.

Given the maturation of cryptocurrency into an asset class, this review shows that cryptocurrency should be studied the same way we study any asset. The right way to understand it is exactly through traditional finance tools---empirical asset pricing. The seven stylized facts we present summarize the key takeaways from this view.


\section*{DISCLOSURE STATEMENT}
The authors are not aware of any affiliations, memberships, funding, or financial holdings that might be perceived as affecting the objectivity of this review. This review is for academic purposes only and should not be construed as investment or legal advice. 




\newpage

\begin{singlespace}
\bibliographystyle{ar-style1}
\bibliography{survey}

\end{singlespace}


\end{document}